# Editorial for the Bibliometric-enhanced Information Retrieval Workshop at ECIR 2014

Philipp Mayr, Philipp Schaer, Andrea Scharnhorst, Peter Mutschke

## 1 Introduction

This first "Bibliometric-enhanced Information Retrieval" (BIR 2014) workshop[1] aims to engage with the IR community about possible links to bibliometrics and scholarly communication [6]. Bibliometric techniques are not yet widely used to enhance retrieval processes in digital libraries, although they offer value-added effects for users. To give an example, recent approaches have shown the possibilities of alternative ranking methods based on citation analysis leading to an enhanced IR. In this workshop we will explore how statistical modelling of scholarship, such as Bradfordizing or network analysis of co-authorship network, can improve retrieval services for specific communities, as well as for large, cross-domain collections. This workshop aims to raise awareness of the missing link between information retrieval (IR) and bibliometrics / scientometrics and to create a common ground for the incorporation of bibliometric-enhanced services into retrieval at the digital library interface. Our interests include information retrieval, information seeking, science modelling, network analysis, and digital libraries. The goal is to apply insights from bibliometrics, scientometrics, and informetrics to concrete practical problems of information retrieval and browsing.

Retrieval evaluations have shown that simple text-based retrieval methods scale up well but do not progress. Traditional retrieval has reached a high level in terms of measures like precision and recall, but scientists and scholars still face challenges present since the early days of digital libraries: mismatches between search terms and indexing terms, overload from result sets that are too large and complex, and the drawbacks of text-based relevance rankings. Such analyses have revealed not only the fundamental laws of Bradford and Lotka, but also network structures and dynamic mechanisms in scientific production. Statistical models of scholarly activities are increasingly used to evaluate specialties, to forecast and discover research trends, and to shape science policy. Their use as tools in navigating scientific information in public digital libraries is a promising but still relatively new development. We will explore how statistical modelling of scholarship can improve retrieval services for specific communities, as well as for large, cross-domain collections. Some of these techniques are already used in working systems but not well integrated in larger scholarly IR environments.

The availability of new IR test collections that contain citation and bibliographic information like the iSearch collection[2] could deliver enough ground to interest (again) the IR community in these kind of bibliographic systems. The long-term research goal is to develop and evaluate new approaches based on informetrics and bibliometrics.

---
[1] http://www.gesis.org/en/events/conferences/ecirworkshop2014/
[2] http://itlab.dbit.dk/~isearch/

The aim of this workshop is to bring together researchers from different domains, such as information retrieval, information seeking, science modelling, bibliometrics, scientometrics, network analysis, and digital libraries to move toward a deeper understanding of this research challenge. In the following we will outline the six papers of the workshop in the sequence of presentation.

## 2    Overview of the papers

Since bibliographic studies enabled the systematic study of citations, researchers have debated about the meaning of citations. The analysis of citations has revealed meaningful traces of knowledge diffusion in scholarly communication based on large scale analysis. This does not take away that for every reference made in a text, the reason for such a reference can be very different. It can be a reference to a body of work fundamental for the argument made in this paper, or indicating other related work with which this paper engages complementary, continuing or debating. Linguistic analysis of the context (the textual neighborhood) of a citation has been conducted to determine the sentiment of a citation. The paper of Bertin and Atanassova [2] belongs to those studies, which try to further unravel the riddle of meaning of citations. The authors analyse the word use in standard parts of articles - such as Introduction, Methods, Results and Discussion, and reveal interesting distributions of the use of verbs for those sections. The authors propose to use this work in future citation classifier, which in the long-term might be implemented also in citation-based information retrieval.

Nees Jan van Eck and Ludo Waltman [4] consider the problem of scientific literature search, and suggest that citation relations between publications can be a helpful instrument in the systematic retrieval process of scientific literature. They introduce a new software tool called CitNetExplorer that can be used for citation-based scientific literature retrieval. To demonstrate the use of CitNetExplorer, they employ the tool to identify publications dealing with the topic of "community detection in networks". They argue that their approach can be especially helpful in situations in which one needs a comprehensive overview of the literature on a certain research topic, for instance in the preparation of a review article.

Muhammad Kamran Abbasi and Ingo Frommholz [1] investigate the benefit of combining polyrepresentation with document clustering. The goal is to provide the search process by highly ranked polyrepresentative clusters. The principle of polyrepresentation in IR can be generally described as the increase of a document's relevancy if multiple representations are pointing to it. Given this, the authors argue that from user perspective it seems more suitable to present clusters of documents relevant to the same representation instead of presenting ranked lists of search results. The approach proposed therefore is to provide the user with a ranked list of documents appearing in the "best" cluster first, i.e. the cluster of documents providing the most cognitive overlap of different representations. The authors applied clustering to information need as well as to document-based polyrepresentation. The evaluation of the model on the basis of the iSearch collection shows some potential of the approach to improve retrieval quality, but also some dependency from the number of relevant documents.

Haozhen Zhao and Xiaohua Hu [7] explore the effect of including citation and co-citation information as document prior probabilities for relevancy on retrieval quality. As document priors a paper's citation count, its PageRank and its co-citation cluster is used. The paper provides an evaluation of the approach on the basis of the iSearch collection, however indicating a limited effect of applying document priors based on citation counts, PageRank and co-citation clusters of retrieval performance. The authors conclude that using document priors in a more query dependent manner and combining citation features with content features might lead to a greater effect.

Zeljko Carevic and Philipp Schaer [3] examined the iSearch test collection and the available citation information included in this collection. Unlike iSearch common IR test collections don't included all available information to do proper evaluations in the field of citation-based rankings. The main goal of this work is to learn about the connection between citation-based and topical relevance rankings and the suitability of iSearch to work on this task. The paper at hand is a pretest for this overall research question and analyses the dataset and it's suitability for citation analysis. Furthermore they investigated on co-citation recommendations based on topical relevant seed documents.

Kris Jack, Pablo López-García, Maya Hristakeva and Roman Kern [5] present a work on how to increase the number of citations to support claims in Wikipedia. They analyse the distribution of more than 9 million citations in Wikipedia and found that more than 400,000 times an explicit marker for a needed citation is present. To overcome this situation they propose different techniques based on journal productivity (Bradfordizing) and popularity (number of readers in Mendeley) to implement a citation recommending system. The evaluation is carried out using the Mendeley corpus with 100 million documents and 10 topics. Although this paper is just a case study it can be clearly seen that a normal keyword-based search engine like Google Scholar is not sufficient to be used to provide citation recommendation for Wikipedia articles and that altmetrics like readership information can improve retrieval and recommendation performance.

## 3   Outlook

After the ISSI workshop "Combining Bibliometrics and Information Retrieval"[3] we aimed with the BIR workshop for a dissemination strategy oriented towards core-IR which is the reason why we located this workshop at ECIR. The variety of papers we received and the small subset we could accept for this workshop show the different ways of combining bibliometrics and IR and show the mutual benefits the two disci-plines can offer each other. We hope to bring both disciplines more closer together and start a sequence of explorations, visions, results documented in scholarly discourse, and set up new material for a sustainable bridge between bibliometrics and IR.

---

[3] http://www.gesis.org/en/events/conferences/issiworkshop2013/